\documentclass{aa}
\usepackage{psfig}

\def\1728{4U\thinspace1728-34}

\def\comment#1{{}}

\begin{document}

\bigskip \bigskip

\thesaurus{06 (02.01.2;  08.09.2;
               08.14.1; 13.25.5)}

\title{Detection of a broad iron emission line and
sub-millisecond quasiperiodic oscillations from the type I
X-ray burster \1728 in a high state}

\author{S.~Piraino,$^{(1)}$ A.~Santangelo,$^{(1)}$
P.~Kaaret$^{(2)}$}

\authorrunning{S.~Piraino et al.}

\titlerunning{BeppoSAX Observations of 4U\,1728-34}

\offprints{Philip Kaaret, email: pkaaret@cfa.harvard.edu}

\institute{
(1) {\it IFCAI/CNR, Via Ugo La Malfa 153, 90146 Palermo, Italy} \\
(2) {\it Harvard-Smithsonian Center for Astrophysics, 60 Garden St., 
     Cambridge, MA 02138, USA}\\ }

\date{Received; accepted}

\maketitle

\begin{abstract}

We report results from simultaneous RossiXTE and BeppoSAX
observations of the neutron-star x-ray binary and type I
X--ray burster 4U\,1728-34.  The source was found in a high
luminosity state, $L_X \sim 0.1 L_{\rm Edd}$, and
quasiperiodic oscillations were detected at $1284 \pm 6$~Hz,
the highest frequency ever observed from this source.  The
x-ray spectrum shows a broad, $\rm FWHM \sim 0.8$~keV, iron
$K_{\alpha}$ fluorescence line.  We discuss interpretations
of the broad line and the quasiperiodic oscillations and how
future simultaneous spectral and timing observations can be
used to test these interpretations and, potentially, to
estimate the mass of the compact object.

\end{abstract}


\section{Introduction}

The behavior of accretion flows around compact objects in
x-ray binaries is of great interest for probing strong
gravitational fields and for understanding the nature of
compact objects.  Two tools used in this pursuit are x-ray
spectroscopy and x-ray timing.  Both have proved useful, but
ambiguities remain in the interpretation of either
spectroscopic or timing observations.  Combination of
simultaneous timing and spectroscopic data may place stronger
constraints on the properties of the accretion flow and on
the compact object.

Here, we combine the timing capabilities of the Rossi X-ray
Timing Explorer (RXTE; \cite{rxte}) with the wide spectral
range and good spectral resolution of BeppoSAX
(\cite{boella97}) in an observation of the neutron-star
low-mass x-ray binary (NS-LMXB) and type I x-ray burster
\1728 made simultaneously with the two observatories.  
Discovered by Uhuru (\cite{form78}), \1728 is a well known
persistent atoll source that shows frequent type I X-ray
bursts (\cite{basinska84}).   The source shows line emission
(\cite{white86}), kilohertz quasiperiodic oscillations (QPOs)
in the persistent emission, and oscillations near 363~Hz in
x-ray bursts (\cite{stroh96}).  

Our observations, as described in \S 2, found the source in
an unusually luminous state.  The timing analysis, including
detection of quasiperiodic oscillations, is described in \S
3, and the spectral analysis, including detection of Fe line
emission, in \S 4.  We interpret these results in \S 5 and
conclude, in \S 6, with a suggestion of a means via which
future observations could be used to test the interpretation.

\section{Observations and Source State}

\1728 was jointly observed by BeppoSAX from 1999 August 19,
01:57:29 UT to August 20 04:54:32 UT for a total on--source
observing time of 51.3~ks and by RXTE from 1999 August 19
03:49:57 UT to August 20 11:26:15 UT for a total good
observing time of 48~ks.  Several x-ray bursts were detected
during these observations.  As we are interested here in the
persistent emission, intervals of the observations
surrounding the bursts were removed.

\begin{figure}[th]
\centerline{\psfig{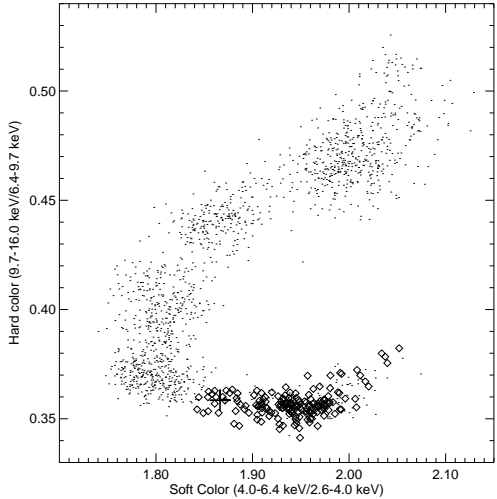}}
\caption[]{A color-color diagram for \1728.  The points
represent 256~s intervals of data from previous RXTE
observations.  The diamonds represent intervals from the
observation analyzed here.  The cross indicates the colors
during the interval when the QPO (see text) was detected.}
\label{fig:cc} \end{figure}

The RXTE data were used to determine the source state, i.e.\
location in the color-color diagram, and to study the timing
behavior.  We extracted x-ray colors in 256~s intervals from
our observation and also, for comparison, from a large data
set extracted from the RXTE archive, see Fig.~\ref{fig:cc}. 
We selected energy bands of 2.6--4--6.4--9.7--16~keV to
define the colors. The lower limit of our energy range is
somewhat higher than typically used previously
(\cite{mendez99}) because recent gain changes in the
Proportional Counter Array (PCA; \cite{zhang93}) do not allow
reliable extraction of flux below 2.6~keV. 

Source position within the color-color diagram changes
continuously along a one-dimensional curve embedded in the
two-dimensional diagram and position along the curve is
generally thought to be an indicator of mass accretion rate
within the system (\cite{hasinger89}; \cite{schulz89}).  For
the colors chosen in Fig.~\ref{fig:cc}, mass accretion rate
increases as the curve is traversed counterclockwise.  From
the position of the source in the color-color diagram, we
infer that the source was in a state of high mass accretion
rate during the observations analyzed here.

\section{Timing behavior}

We searched for QPOs in the range 400--2000~Hz in intervals
of continuous observation, summing 2~s power spectra
calculated from $122 \, \mu$s time resolution PCA event data
within each interval. We first conducted a search over 256~s
intervals using all events, but did not find any significant
QPO peaks.  Previous QPO searches using relatively short
intervals in observations of \1728 have shown that the QPO
amplitudes drop below the threshold of detection at high
inferred mass accretion rates (\cite{mendez99}), such as
found during our observation. Thus, this null result was not
unexpected.

\begin{figure}[th]
\centerline{\psfig{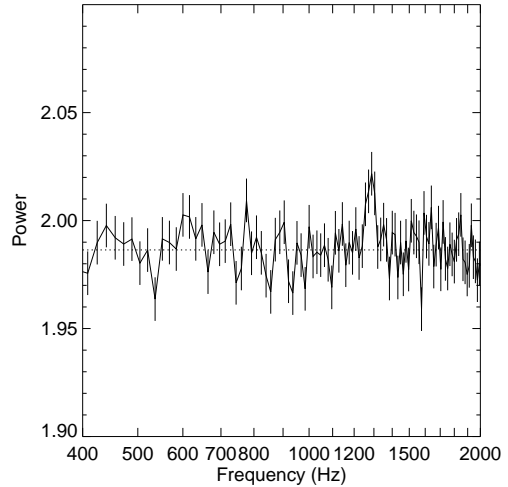}}
\caption[]{Power spectrum for the 2460~s interval containing
the QPO at $1284 \pm 6$~Hz.  The powers are Leahy normalized
and have been rebinned for presentation.  The dotted line
shows the calculated Poisson noise level.}
\label{fig:powspec} \end{figure}

From analysis of archival observations at the highest
inferred mass accretion rates where kHz QPOs were detected,
we found that selection of events with energies above 4~keV
optimized the signal to noise ratio for detection of kHz
QPOs.  We used this energy selection criterion to perform a
new kHz QPO search in the current data.  To maximize the
statistics available in each power spectrum, we used full
continuous observation intervals with typical durations near
3~ks.  The only detection of a significant kHz QPO peak in
our search range (400--2000~Hz) in our data is of a QPO peak
at $1284 \pm 6$~Hz in an interval with a duration of 2460~s,
see Fig.~\ref{fig:powspec}.  The QPO peak has a width of $32
\pm 12 \rm \, Hz$ and an rms amplitude of $3.9\% \pm 0.5\%$
(above 4~keV).   Allowing for 75 trials, we estimate the
probability of chance occurrence to be $1.5 \times 10^{-5}$,
corresponding to a $4.3 \sigma$ detection.

The QPO frequency is the highest ever reported for \1728.  
In \1728, the frequency of kHz QPOs is well correlated with
position in the color-color diagram with higher frequencies
corresponding to higher inferred mass accretion rates
(\cite{mendez99}).  The x-ray colors during the interval when
the QPO is detected (1.866, 0.358) are indicated by the cross
in Fig.~\ref{fig:cc} and correspond to a higher inferred mass
accretion rate than any previous kHz QPO detection in this
source.

\begin{figure}[th]
\centerline{\psfig{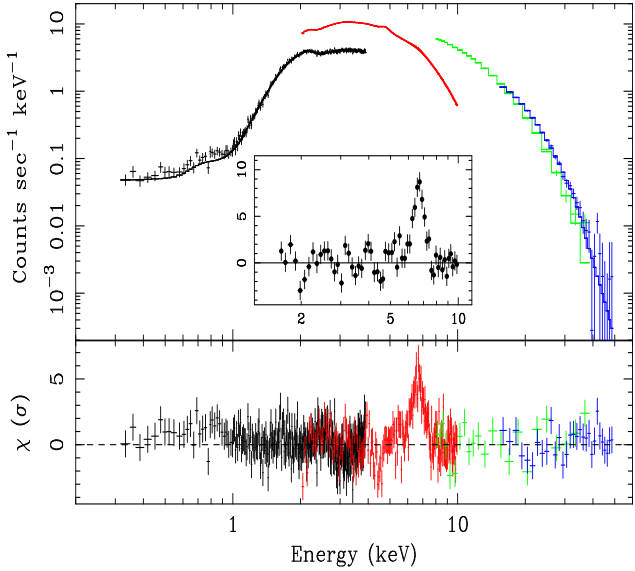}}
\caption[]{The 0.1-50~keV spectrum of \1728 observed by
BeppoSAX is shown together with the residuals in the entire
band, in unit of standard deviations, when the best fit
continuum is applied in the whole band except the 4--8~keV
energy range.  The inset shows the residuals of the MECS data
rebinned to better display the profile of the observed Fe
$K_{\alpha}$ feature.} \label{fig:spec} \end{figure}

\section{Spectral behavior}

Broad band energy spectra of the source were obtained
combining data from the four BeppoSAX Narrow Field
Instruments (NFIs): the Low Energy Concentrator Spectrometer
(LECS; \cite{parmar97}) for 0.3--4~keV, the Medium Energy
Concentrator Spectrometer (MECS; \cite{boella97}) for
1.8--10~keV, the High Pressure Gas Scintillation Proportional
Counter (HPGSPC; \cite{manzo97}) for 8--40~keV, and the
Phoswich Detection System (PDS; \cite{frontera97}) for
15--50~keV.  LECS and MECS data were extracted in circular
regions centered on the source position using radii of 8' and
4' respectively, containing 95\% of the source flux.  An
image analysis of the BeppoSAX data revealed no other
sources.  There is a source in the WGACAT (\cite{wgacat})
$5\arcmin$ away, but its flux is a factor of 100 lower and
would not significantly contaminate the spectra.  The spectra
have been rebinned to have at least 30 counts per channel,
and the HPGSPC and PDS spectra were grouped using a
logarithmic grid.  Standard normalization factors have been
included to account for the mismatch in the BeppoSAX
instruments absolute flux calibration.  We found that spectra
extracted for different intervals within the observations
gave spectral parameters consistent within errors.  Thus, we
report spectral fits only for the sum over the entire
BeppoSAX observation.

Different models were used to fit the broad band continuum.
The one that gave the best fit contained the following
components: 1) a black body at soft energies (\cite{white88})
described by a temperature ($kT_{\rm bb}$) and flux ($F_{\rm
bb}$), 2) a Comptonized component (COMPTT, \cite{titar94})
for the hard energy part described by the temperature of
injected photons ($kT_{\gamma}$), the electron temperature
($kT_{e}$), and optical depth ($\tau$), 3) photoelectric
absorption at low energy with a column density ($N_{\rm
H}$).  However, even this model alone was clearly rejected,
$\chi^2_{\nu} \sim 1.65$, and strong residuals were present
between 4 and 8~keV indicative of Fe line emission.  In order
to find the best fit continuum parameters, we fitted the
continuum excluding the data between 4 and 8~keV.  In
Fig.~\ref{fig:spec}, the spectrum together with residuals
from this fit are shown.  The line emission is clearly
broader than the instrumental response ($\sigma = 0.20 \rm \,
keV$ at 6~keV); using a single narrow line in place of the
broad line increases the $\chi^2$ by 30.  The line was
modeled adding a simple Gaussian profile with centroid
($E_{\rm G}$), width ($\sigma$), and equivalent width
($EW$).  The best fit had $\rm \chi^2/DoF =  653/516$ and the
best fit parameters are reported in Table~\ref{tab:spec}. 
The 0.2-50~keV flux was $6.4 \times 10^{-9} \rm  \, erg \,
cm^{-2} \, s^{-1}$ and the unabsorbed flux was $8.6 \times
10^{-9} \rm  \, erg \, cm^{-2} \, s^{-1}$.  We also searched
for an absorption edge in the energy range 7-10~keV.  No
statistically significant edge was detected and we place an
upper limit of 0.19 (95\% confidence) on the optical depth of
any absorption edge in the 7-10~keV band.

\begin{table}[t]
\caption[]{Spectral parameters  of \1728. } \label{tab:spec}
\begin{center}
\begin{tabular}{l|l}
\noalign{\smallskip}
\textbf{Parameter}                                       & \textbf{Value} \\
 \hline
 $kT_{\gamma} \rm [keV]$                                 & $1.16 \pm 0.03$ \\
 $kT_{e} \rm [keV]$                                      & $3.16 \pm 0.03$ \\
 $\tau$                                                  & $11.4 \pm 0.2$ \\
 $kT_{\rm bb} \rm [keV]$                                 & $0.57 \pm 0.01$ \\
 $F_{\rm bb} \rm [10^{-9} \, erg \, cm^{-2} \, s^{-1}]$  & $2.18 \pm 0.05$ \\
 $E_{\rm G} \rm [keV]$                                   & $6.72 \pm 0.05$ \\
 $\sigma \rm [keV]$                                      & $0.34 \pm 0.08$ \\
 $EW \rm [eV]$                                           & $52 \pm 9$ \\
 $N_{\rm H} \rm [10^{22} cm^{-2}$]                       & $2.73 \pm 0.05$ \\
\noalign{\smallskip}
\end{tabular}
\end{center}
{\small \sc Note} \small--- All quoted errors represent
$90\%$ confidence level for a single parameter.
\end{table}

\section{Discussion}

The line could arise from the neutron star itself, the
accretion disk, or a corona.  The high Compton optical depth
of the continuum emission likely excludes line production on
the neutron star surface as the line would be down-scattered
and significantly reduced in intensity.  However, line
production on the stellar surface cannot be completely
excluded if the Comptonizing region is small compared to the
stellar radius or comparable to the stellar radius and highly
non-uniform.

Production of the line in an accretion disk corona (ADC)
should lead to an Fe absorption edge in addition to the
emission line.  Broadening of a single narrow line via
Compton scattering to the width observed would require an
optical depth greater than 2.6, in strong contrast to our
upper limit of 0.19.  Unless the ADC is very highly ionized,
suppressing the Fe edge, the low upper limit obtained on the
optical depth would argue against significant broadening due
to Comptonization.  However, the observed broad feature could
be a blend of lines originating from an ADC and broadening
due to rotation is possible.

Origin of the line in the accretion disk was disregarded in
many past studies of broad iron emission features from LMXBs
(e.g. \cite{white85}) due to the high ionization expected in
the inner parts of the disk.  However, the luminosity of
\1728 is $L_{X} \sim 0.1 L_{\rm Edd}$ significantly lower
than $L_{X} \sim L_{\rm Edd}$ seen from the brightest LMXBs. 
Using a thin disk model (\cite{shakura73}) in the radiation
pressure dominated regime and taking a distance to \1728 of
4.2~kpc (\cite{vp78}) leading to a luminosity $L_{X} \sim 0.1
L_{\rm Edd}$, a viscosity parameter in the range $\alpha =
0.2-0.5$, and assuming illumination from a point source
located at the center of the neutron star, we find that the
peak ionization, found near a radius of 21~km, is in the
range $\xi = 500 - 1200 \rm \, erg \, cm \, s^{-1}$.  This is
only a rough estimate since the peak ionization depends on
the assumed distance, viscosity, disk model, and source
geometry.  However, significant iron line flux is expected in
this ionization range (\cite{matt96}).  Our measured line
profile is not inconsistent with relativistically broadened
line emission from highly ionized iron (Fe {\sc XXV} to Fe
{\sc XXVI}).  To avoid Compton down-scattering similar to
that expected to render the line undetectable if emitted from
the stellar surface, the Comptonization region must be quite
compact.  Recent results on energy lags in kHz QPOs in
NS-LMXBS favor Comptonization regions with sizes no larger
than 10~km (\cite{kaaret99}).  Thus, the accretion disk
appears to be a plausible site of origin for the observed
line emission.

\section{Summary and Outlook}

We have obtained simultaneous measurements of the x-ray
spectrum and the high frequency timing properties of the
low-mass x-ray binary \1728.  We report detection of the
highest frequency oscillations, at $1284 \pm 6$~Hz, yet
observed from this source.  This is the second highest kHz
QPO frequency reported and the highest with a statistical
significance above $4 \sigma$ (\cite{vanStra}).  Detection of
this high frequency may be used to place constraints on the
mass-radius relation of the neutron star in \1728
(\cite{miller98}).

We also report detection of a broad emission feature.  We
suggest that the emission feature may be a relativistically
broaden line emitted from the ionized inner accretion disk
(\cite{smale93}).  However, the line could, instead, arise
from an accretion disk corona, and new observations with much
better spectral resolution are called for to constrain the
ionization state of the emitting matter and help determine
the origin of the line.

To conclude, we offer a few speculations on information that
could obtained from future simultaneous timing and spectral
observations if the Fe line does arise from the inner
accretion disk.  Most of the leading models of kHz QPOs in
NS-LMXBs identify one kHz QPO frequency with the Keplerian
orbital frequency at the inner edge of an accretion disk
(\cite{miller98, psaltis00, stella99, titar99}).  A key test
of this interpretation would be to find evidence for changes
in the disk properties measured via other, i.e. spectral,
means correlated with changes in the QPOs.  If an Fe line
from the inner accretion disk is observable, then the Fe line
profile, via relativistic broadening, should provide
information on the inner radius of the disk.  The QPO
frequency is also determined by the inner radius of the disk
and the radius and frequency should be related via Kepler's
third law, $\nu^2 \propto r^{-3}$.  Detection of such a
correlation would be a striking confirmation of the
interpretation of the kHz QPO frequency as a Keplerian
orbital frequency.  Furthermore, if this interpretation is
correct, simultaneous knowledge of an orbital radius and the
associated orbital frequency would lead to a measurement of
the compact object mass.  Accurate mass determinations are
available for binary radio pulsars, but not for accreting
neutron stars (\cite{vp98}) which are of interest because
they have undergone significant accretion and, thus, should
probe the allowed mass range of neutron stars.

\begin{acknowledgements}

We thank Evan Smith and Donatella Ricci for their efforts in
coordinating the observations and the referee, Norbert
Schulz, for comments which improved the paper.  PK and SP
acknowledge support from NASA grants NAG5-7405 and NAG5-8408.
SP acknowledges support from a CNR short-term mobility grant.

\end{acknowledgements}

\end{document}